# A Cryogenic Tune and Match Circuit for Magnetic Resonance Microscopy at 15.2T

*Benjamin M Hardy[1,2], Gary Drake[2], Shuyang Chai[2], Bibek Dhakal[1,2], Jonathan B Martin[2,4], Junzhong Xu[1,2,3,4], Mark D Does[2,4], Adam W Anderson[2,3,4], Xinqiang Yan[2,3], John C Gore[1,2,3,4]*

[1]Department of Physics and Astronomy, Vanderbilt University, Nashville, TN 37232, USA
[2]Institute of Imaging Science, Vanderbilt University Medical Center, Nashville, TN 37232, USA
[3]Department of Radiology and Radiological Sciences, Vanderbilt University Medical Center, Nashville, TN 37232, USA
[4]Department of Biomedical Engineering, Vanderbilt University, Nashville, TN 37232, USA

\* **Corresponding author**: Address: Vanderbilt University, Institute of Imaging Science, 1161 21st Avenue South, AA 1112 MCN, Nashville, TN 37232-2310, United States.
E-mail: Benjamin.hardy@remcom.com (Benjamin M. Hardy)

Abstract

Background and Significance:

Achievable signal to noise ratios (SNR) in magnetic resonance microscopy images are limited by acquisition times and the decreasing number of spins in smaller voxels. A common method of enhancing SNR is to cool the RF receiver coil. Significant SNR gains are realized only when the Johnson noise generated within the RF hardware is large compared to the electromagnetic noise produced by the sample. Cryogenic cooling of imaging probes is in common use in high field systems, but it is difficult to insulate a sample from the extreme temperatures involved and in practice imaging cryoprobes have been limited to surface or partial volume designs only. In order to be able to use a solenoid in which the windings were not cooled and in close proximity to the sample, we designed a chamber to cool only the tune and match circuitry and show experimentally it is possible to achieve much of the theoretically available SNR gain.

Methods: A microcoil circuit consisting of two tuning capacitors, one fixed capacitor, and SMB coaxial cable was designed to resonate at 650 MHz for imaging on a Bruker 15.2 T scanner. Sample noise increases with the sample diameter, so surface loops and solenoids of varying diameters were tested on the bench to determine the largest diameter coil that demonstrated significant SNR gains from cooling. A liquid $N_2$ cryochamber was designed to cool the tune and match circuit, coaxial cable, and connectors, while leaving the RF coil in ambient air. As the cryochamber was filled with liquid $N_2$, quality factors were measured on the bench while monitoring the coil's surface temperature. Improvements of SNR on images of ionic solutions were demonstrated via cooling the tune and match circuit in the magnet bore.

Results: At 650 MHz, loops and solenoids < 3 mm in diameter showed significant improvements in quality factor on the bench. The resistance of the variable capacitors and the coaxial cable were measured to be 45% and 32% of room temperature values near the Larmor frequency. Images obtained with a 2 turn, 3 mm diameter loop with the matching circuit at room temperature and then cooled with liquid nitrogen demonstrated SNR improvements of a factor of 2.

Conclusions By cooling the tune and match circuit and leaving the surface loop in ambient air, SNR was improved by a factor of 2. The results are significant because it allows for more space to insulate the sample from extreme temperatures.

## Introduction

Magnetic resonance microscopy (MRM) is potentially capable of resolving microstructural features in biological tissues [1–5]. MRM images may exploit various contrast mechanisms such as relaxation rates and diffusion, and are able to probe structures relatively deep within volumetric tissue compared to e.g. optical imaging. However, as voxels shrink in size, the number of excitable spins quickly diminishes, so signals become weaker. Furthermore, diffusion attenuates signals and blurs the point spread function, ultimately limiting the achievable spatial resolution [6–9]. Several methods have been proposed to overcome the signal-to-noise (SNR) and resolution limits of MRM. Increased sensitivity may be achieved by using smaller RF coils typically referred to as microcoils [10] as SNR is inversely proportional to the coil diameter. The smallest reported microcoils are usually surface loops or solenoids 100 - 500 µm in diameter. The highest resolution MRM images reported to date had resolution ≈ 3 µm [11,12], with the highest being 2.7 µm isotropic with a liquid helium cooled solenoid [13].

Cooling circuitry and hardware improves circuit efficiencies for many electronic applications and industries [14,15]. Cooling reduces the Johnson noise within resistive components that is the result of thermal agitation of electrons. Hoult and Richards analyzed the strength of the noise generated within RF coils in NMR [16] as a guide for design and performance metrics. As such, cooling is common in high resolution multi-nuclear NMR. Styles and coworkers were some of the first to cool the preamp and RF coil with liquid helium, and improved SNR by a factor of 9 for $C^{13}$ spectroscopy at 1T for small liquid samples [17]. The overall SNR improvement depends on the operating frequency and the size of the sample thus increases in SNR for cooled systems have ranged from 1.5 to 9 [18–21].

Whether surface loop or solenoid, the inductor is often cooled via immersion or contact with a cold head. Thus, when the sample is very close to the coil, as is desired, a key challenge is insulating the sample from the extreme temperatures of the windings to avoid freezing the sample.

There have been no reported cases of immersion cooling of only the coaxial cable and tune and match circuit while keeping the inductor in open air. Thus, we aimed to leave the inductor in ambient air with the sample while cooling the tune and match circuit to liquid $N_2$ temperatures. The resistance contributions of the transmission line elements were investigated at varying temperatures. SNR gains are demonstrated by cooling the tune and match circuit in a custom-built double walled $N_2$ chamber. The chamber is designed such that the inductor and sample are in open air and only the tune and match components and coax are immersed in $N_2$. To the author's knowledge, this is the first use of a cryogenically cooled circuit while keeping the inductor of the RF coil outside of the cryogenic environment in ambient air.

# Theory

## 2.1 Resistance contributions

The noise in an MR acquisition depends on several factors [22]. We can define the noise of the receiver chain as the temperature of a resistance that would provide the same noise power. Then

$$Noise_{MR} = \sqrt{4k_b BW T_{total} R_{total}} \quad (1)$$

Where $k_b$ is boltzmann's constant, $T_{total}$ is the temperature of the system, BW is the bandwidth of the receiver, and $R_{total}$ is the combined series resistances of the hardware and the sample considered as a lossy material coupled to the coil i.e.

$$R_{total} = R_{sample} + R_{circuit} \quad (2)$$

The relative contributions of the sample and coil have been well characterized and studied [16,23]. Darrasse and Ginefri provide a thorough discussion surrounding the relationship between frequency, sensitivity, coil size, and temperature [24]. Their estimate of the coil and sample noise dominated regimes uses a surface loop and an infinite conductive sample, which underestimates the noise contributions of the RF receiver chain. By including resistance of other lumped components including the variable capacitors, with a generous Q of 1000, and a modest lead length of 5 mm, the coil noise dominated region is pushed out towards increasing coil diameters (figure 1).

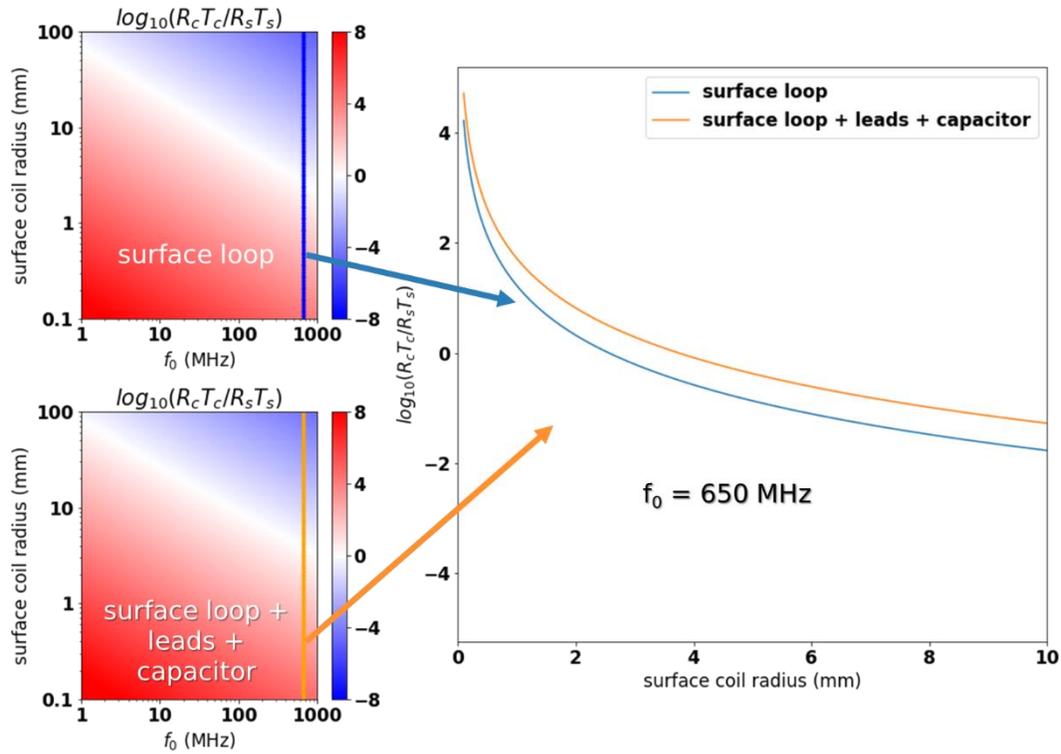

Figure 1: By considering series resistance contributions of the capacitor (Q=1000) and necessary leads to the inductive loop, larger coil sizes are included in the coil noise dominated region. The capacitor and lead losses are not trivial in the RF receiver chain for microcoils at 650 MHz.

The two main contributions of noise from the sample are magnetic and dielectric losses where

$$R_{sample} = R_{magnetic} + R_{dielectric} \quad (3)$$

Minard and Wind [25,26] considered sample losses for microcoils and provide specific and experimentally verified descriptions of the losses from the sample. The following passages follow their work closely with some further simplification and extrapolation to higher frequencies.

### 2.1.1 Sample Magnetic Losses

In a conducting sample, magnetic losses are caused by induced eddy currents by the rotating magnetic field from the RF coil. For a homogeneous sample of cylindrical shape [23,26], these losses may be written as,

$$R_{magnetic} = \frac{\pi \omega_0^2 \mu_0^2 n^2 A^4 B \sigma}{128(d^2 + B^2)} \quad (4)$$

Where $\omega_0 = 2\pi f_0$, $\mu_0$ is the magnetic permeability of free space, n is the number of turns, A is the sample radius, B is the sample length, and σ is the conductivity of the sample. For microcoils, magnetic losses contribute much less than the dielectric losses (supplementary figure 1).

### 2.1.2 Sample Dielectric Losses

The electric field lines from the coil inevitably pass through the conductive sample, so energy is dissipated in the form of conduction, dielectric relaxation, and dielectric resonance [27,28]. Modelling the dielectric losses can be quite complicated without full-wave simulation so Minard and Wind [26] approximate the resistance by modeling the coil and sample as equivalent circuit models. By combining Minard and Wind's equations 18-20 and 22-26 a simplified expression is obtained for the effective dielectric losses from the sample

$$R_{dielectric} = \frac{\omega_0^3 L^2 C_{stray} \varepsilon'' f_d}{2[\varepsilon''(1-f_d)^2 + f_d^2(\varepsilon'-1)^2 + \varepsilon'^2(1-2f_d) + 2\varepsilon' f_d]} \quad (5)$$

Where L is the inductance of the coil, ε is the complex frequency dependent permittivity of the sample where $\varepsilon(\omega) = \varepsilon' - i\varepsilon''$, and $f_d$ is the experimentally determined dielectric filling fraction of the coil. $C_{stray}$ is the parasitic capacitance of the coil outlined by Medhurst [29], where

$$C_{stray} = d\left(0.1126\frac{l_{coil}}{d} + 0.08 + \frac{0.27}{\sqrt{l_{coil}/d}}\right) 10^{-10} \quad (6)$$

Where $l_{coil}$ is the length of the solenoid or width of surface loop. We follow Minard and Wind's experimentally verified approach and set $f_d = 0.948$ because the coil sizes investigated here are similar. The frequency dependent complex permittivity of ionic samples is well-studied and may be calculated using the following equations.

$$\varepsilon(\omega)' = \varepsilon_\infty + \frac{\varepsilon(0) - \varepsilon_\infty}{1 + \omega^2 \tau^2} \text{ and } \varepsilon(\omega)'' = \frac{\sigma}{\omega \varepsilon_0} + \frac{(\varepsilon(0) - \varepsilon_\infty)\omega\tau}{1 + \omega^2 \tau^2} \quad (7)$$

Where $\varepsilon_\infty$ and $\varepsilon(0)$ the permittivity at very high and low frequencies, τ is the dielectric relaxation time, and $\varepsilon_0$ is the permittivity of free space. For an ionic water solution at 25 °C with σ = 1 S/m, ε(0) = 78.32, $\varepsilon_\infty$ = 5.30, and τ = 8.27 x $10^{-12}$ s [30].

### 2.1.3 Resistance contributions from the circuit

Capacitors, connectors, coaxial cable, the coil, and match circuit all contribute to the total resistance of the probe. Hardware dominates the noise in a microcoil [18], so small changes in the circuit topology could result in significant changes in the SNR of the circuit. The total resistance of the coil may be expressed as a series of lumped resistive elements where

$$R_{circuit} = R_{coax} + R_{connectors} + 2R_{cap} + R_{coil} \quad (8)$$

Where $R_{coax}$ is the resistance of the coaxial cable, $R_{connectors}$ the resistance from cable connectors and leads between elements, $R_{cap}$ is the resistance of a variable capacitor multiplied by 2 (the tune and match capacitor), and $R_{coil}$ the resistance of the solenoid or surface loop. Most of these items have losses characterized by industry standard resonant line methods [31]. For the coaxial cable, textbooks, online calculators, and the product data sheets provide loss per unit length [32] and connectors have added resistance which can usually be found in their respective data sheet.

Minard et al. and Hoult et al. use the skin effect to calculate the resistive contributions from wire for the leads and a solenoid [16,26]. For a solenoid the resistance may be described with the following equation

$$R_{solenoid} = \frac{l\xi}{d_{wire}} \sqrt{\frac{\mu_0 \mu_r \rho f_0}{\pi}} \quad (9)$$

Where l is the total wire length (l = ndπ), $d_{wire}$ is the wire diameter, ξ is the resistance enhancement factor to account for the added eddy currents from adjacent turns, ρ is the resistivity of the conductor (1.72 x $10^{-8}$ Ωm for Cu) and $\mu_r$ is the relative permeability of the conductor ($\mu_r \approx 1$ for Cu). It is interesting to note that ρ is temperature dependent and could be lowered with temperature or a better conductor such as silver [33,34].

The ξ factor is determined by the theoretical predictions of Butterworth [35] and was experimentally verified by Medhurst [29]. The ξ values may be found for varying d/$l_{coil}$ and $d_{wire}$/s ratios in table 1 of Medhurst and Minard et al where *s* is the spacing between turns (s = $l_{coil}$/n). ξ may be calculated by interpolating across the table.

For a surface loop the resistance is given by Darrasse and Ginefri,

$$R_{loop} = \frac{2d\xi n^2}{d_{wire}} \sqrt{\rho \mu_0 \pi f_0} \quad (10)$$

For a capacitor the equivalent series resistance (ESR) may be related to the Q, operating frequency, and capacitance with the following equation.

$$R_{cap} = \frac{1}{2\pi f_0 Q_{cap} C} \quad (11)$$

Where $Q_{cap}$ is the quality factor of the capacitor and C is its capacitance. Most MR RF circuits use at least two variable capacitors for tuning to the Larmor frequency and matching the characteristic impedance ($Z_0$ = 50 Ω).

2.3 Relative contributions considering coil size and temperature

The calculated SNR gains may then be inferred [19] by combining equations 1-13 where

$$SNR_{gains} = \frac{SNR_{N_2}}{SNR_{RT}} \propto \frac{\sqrt{T_{sample,RT}R_{sample,RT} + T_{circuit,RT}R_{circuit,RT}}}{\sqrt{T_{sample,N_2}R_{sample,N_2} + T_{circuit,N_2}R_{circuit,N_2}}} (12)$$

The SNR gains may be further related to the temperature and resistance of each circuit element in equation 9.

$$\frac{SNR_{N_2}}{SNR_{RT}} \propto \frac{\sqrt{T_{sample,RT}R_{sample,RT} + T_{cap,RT}R_{cap,RT} + T_{coax,RT}R_{coax,RT} \cdots}}{\sqrt{T_{sample,N_2}R_{sample,N_2} + T_{cap,N_2}R_{cap,N_2} + T_{coax,N_2}R_{coax,N_2} \cdots}} (13)$$

Predictions for a solenoid and surface loop are plotted in figure 2 using equations 1-13. SNR gains are plotted for both geometries with two scenarios: 1) cooling the full circuit (every element in eq. 8) to liquid nitrogen temperatures 77K and 2) partially cooling the resonant geometry while cooling the rest of the circuit to 77K.

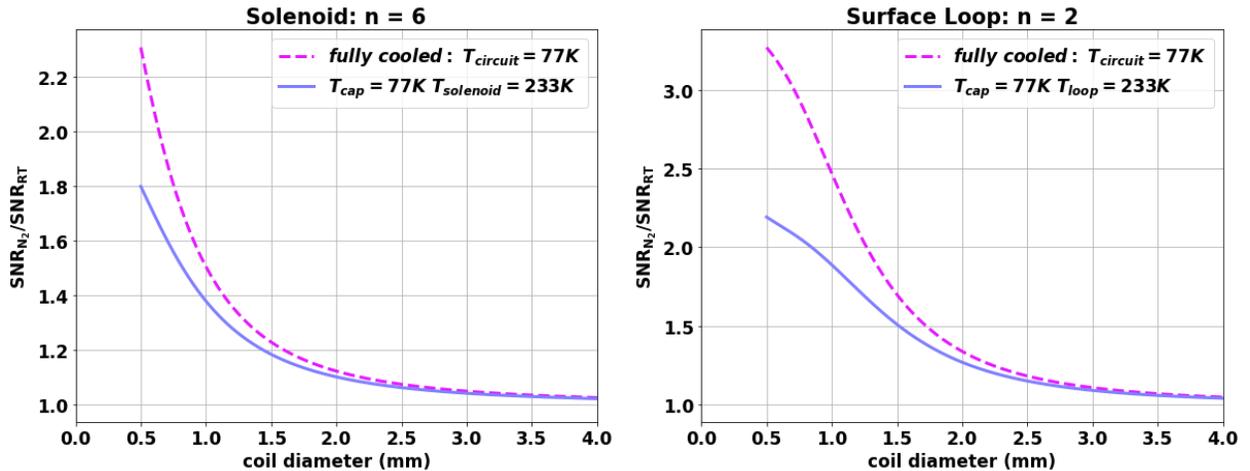

Figure 2: Predictions for two microcoil geometries. The dashed line represents SNR gains from cooling the entirety of the circuit to liquid nitrogen temperatures (77 K). The solid line represents partially cooling the solenoid/loop to 233K and cooling the rest of the circuit to 77 K. The model does not account for the capacitor change in Q with temperature.

In figure 2, the sample had a conductivity of 0.5 S/m and was scaled with the coil size where A = 0.8d and B = 5*$l_{coil}$. The surface loop had a fixed length. The solenoid length scaled with coil diameter where $l_{coil}$ = 1.2d. The quality factor of the capacitor was set to 2000.

All equations and figures from section 2 are available on github at https://github.com/benjhardy/Coil_vs_Sample_Noise_MRI including the coil class that summarizes Minard and Winds publications in a python class for ideal micro-solenoid design choices.

## Methods

### 3.1 The magnet and equipment

All images were acquired with a Bruker (Billerica, MA) 15.2T Biospec imaging spectrometer equipped with max 1 T/m magnetic field gradients (Resonance Research Inc, Billerica, MA). The Avance III console used ParaVision 6.0.1. The bore is 6 cm in diameter within the gradients with 3rd order B0 shims.

All Q measurements were performed with a Keysight (Santa Rosa, CA) ENA Network Analyzer E5080A. All temperature measurements were performed with an Amprobe (Everett, WA) TMD-56 multilogger thermometer and flat-pin, k-type, thermocouple probes.

The small surface loop and circuit trace was milled using an LPKF (Garbsen, Germany) Protomat S103 PCB router. The substrate of the surface loop was Rogers (Chandler, AZ) R3006. Solenoids were wound around Drummond (Broomall, PA) capillaries using copper wire. Air tubular trimmer capacitors were acquired from Passive Plus Incorporated (Huntington, New York) and had a capacitance range of 0.3-10 pF. Temperature of the fomblin (Solvay, Brussels, Belgium) bath was maintained using a Fisherbrand (Waltham, MA) Isotemp Refrigerated/Heated Bath Circulator.

### 3.2 Liquid $N_2$ Cryochamber

The description of the liquid nitrogen cryochamber is given by figure 3 and the details of its assembly are may be found in appendix 1.

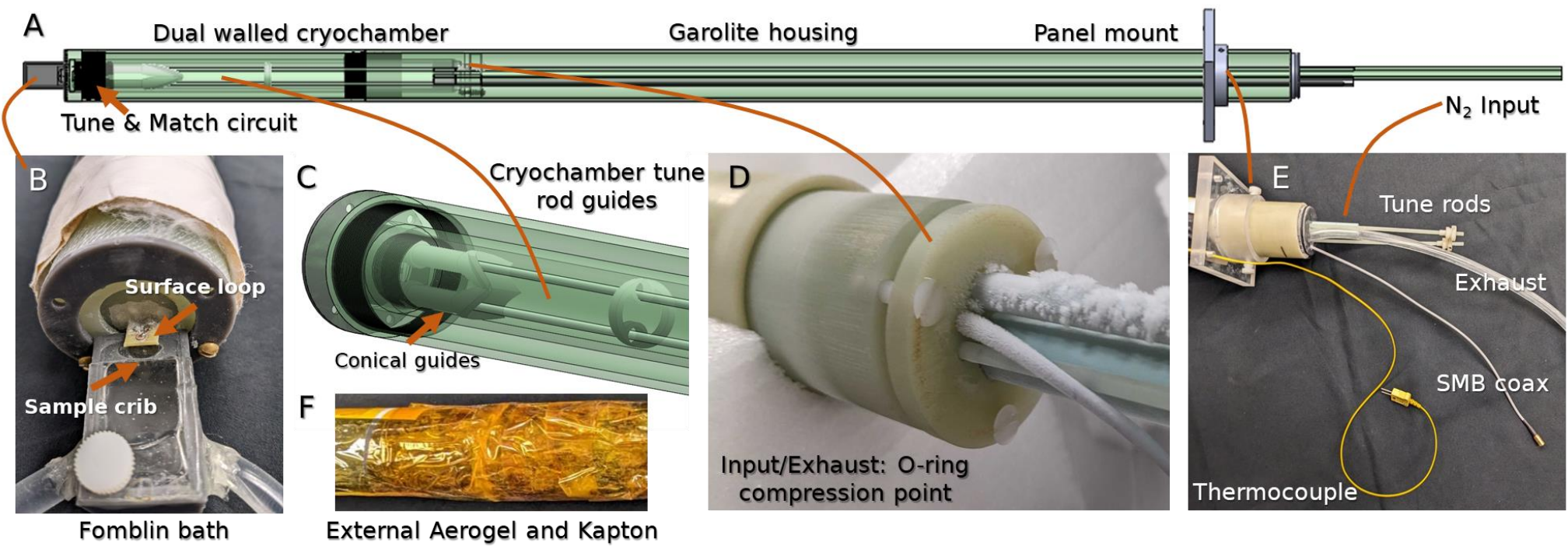

Figure 3: The cryochamber key components include the tune and match circuit and surface loop (A and B), the dual walled cryochamber (A and C), the input/exhaust and O-ring compression point (D), and the garolite housing and insulative material (A and F). The fomblin bath is installed onto the front external cap via brass screws (B). Within the chamber conical guides were 3D printed to thread onto the variable capacitors (C). A floating guide was also placed within the chamber to ensure the rods contacted the conical guide (loop in C). The O-rings were compressed with a garolite mounted onto the back external cap (D). Tune rods, exhaust, SMB coax, and input line flowed through the housing past the magnet panel mount for easy access (E). The surface of the entire assembly was surrounded by Aerogel, heat shrink, and Kapton tape for insulation (F).

### 3.3 Sample crib and Fomblin bath

In some bench tests (see 4.1 and 4.2), the sample would freeze since the coil wires are indirectly cooled via thermal conduction. To maintain sample temperature above 0°C a sample insulation system was designed utilizing a small fomblin bath (figure 4). Temperature controlled water circulated through a plastic tube embedded into the fomblin bath. The setup exploits the existing systems for temperature control in the bore while avoiding any artefacts from the flow of water near the FOV. The fomblin is invisible to the MR experiment thus there is negligible added noise. The sample crib was designed within a small cage so that any air bubbles would be trapped on the outside of the FOV at the surface of the bath. The fomblin bath was covered and sealed using a hot glue and thin microscope glass. The water was run at room temperature for testing within the bore.

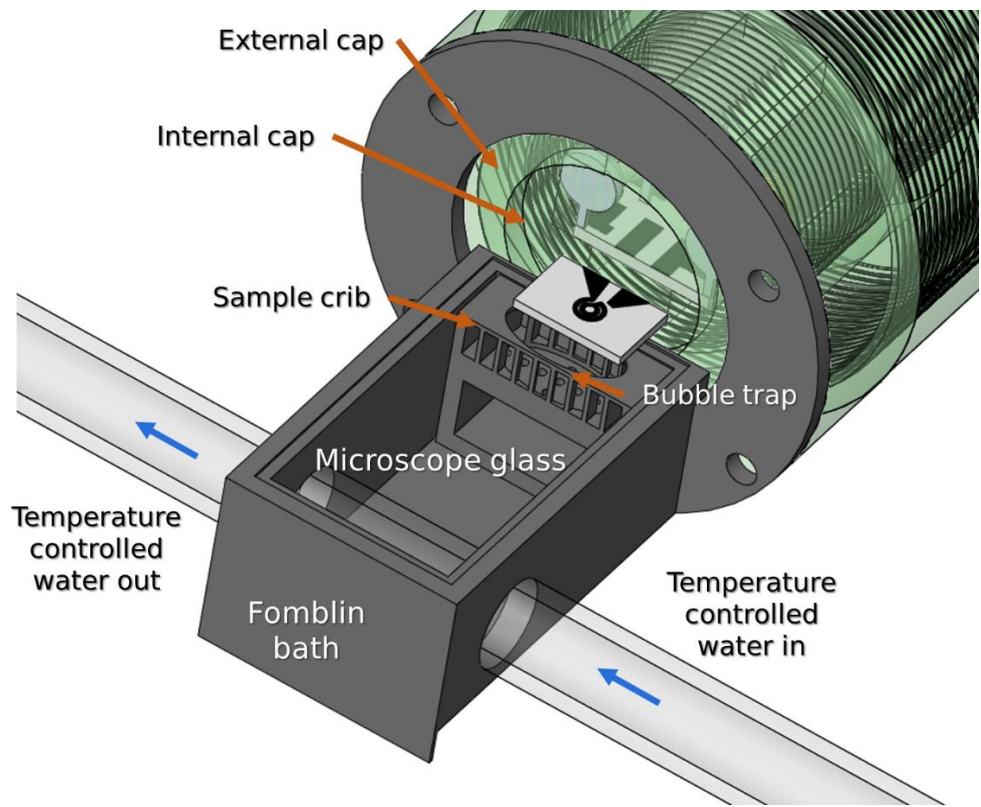

Figure 4: The fomblin bath was designed to insulate the sample from extreme temperatures on the surface of the internal cap and the trace of the surface loop. Temperature controlled water flows through a plastic tube that heats the fomblin and inevitably the sample via convection. The sample is suspended in a crib directly under the surface loop. The crib acts as a bubble trap if bubbles are kept on the opposite side of the sample. A small microscope glass slide was placed over sample crib to minimize the distance between the coil and sample.

### 3.4 Bench Measurements

After an Open-Short-Load (OSL) calibration on the network analyzer, the coil was tuned and matched to resonate at 650 MHz. The $S_{11}$ curve is exported for processing with a python script where the peak and the -3 dB points to the left and right of the resonance are used to determine the bandwidth at half-power or the full-width-half-max (FWHM). The quality factor of the resonating circuit is then calculated with the following equation

$$Q\,(S_{11}) = 2 \frac{f_0}{f_{right}(-3\,dB) - f_{left}(-3\,dB)} \quad (14)$$

Where $f_0$ is the resonance frequency and the factor of 2 accounts for the signal's forward and backward traveling wave. Following appendix 2, equivalent series resistances (ESR) of lumped elements were measured on the bench.

### 3.5 Imaging experiments

To verify that bench tests translated to SNR improvements for MR imaging, a series of images were acquired as the cryochamber cooled down. The following is a description of the experimental setup and the corresponding results. The water circulator for the fomblin bath was set to 22°C. A thermocouple was placed at the observed coldest point of the external insulation on the cryochamber.

After tuning, matching, shimming, and power calibration, sets of FLASH images were acquired at room temperature. The images were 128 x 128 pixels, averaged over 2 acquisitions with 4 slices and 10 repetitions making for a total of 40 images for each set. The FOV was set to 30 x 30 mm with 600 µm slice thickness. TE/TR = 2.6/50ms and the total time for the set was 2 min and 8 s. This same set of 10 images x 4 slices was acquired as the cryochamber cooled down. To avoid damaging the water-cooled gradient coil, cooling of the surface of the cryochamber was limited to ≈ -25°C.

# Results

## 4.1 Q-factor measurements on the Bench without cryochamber

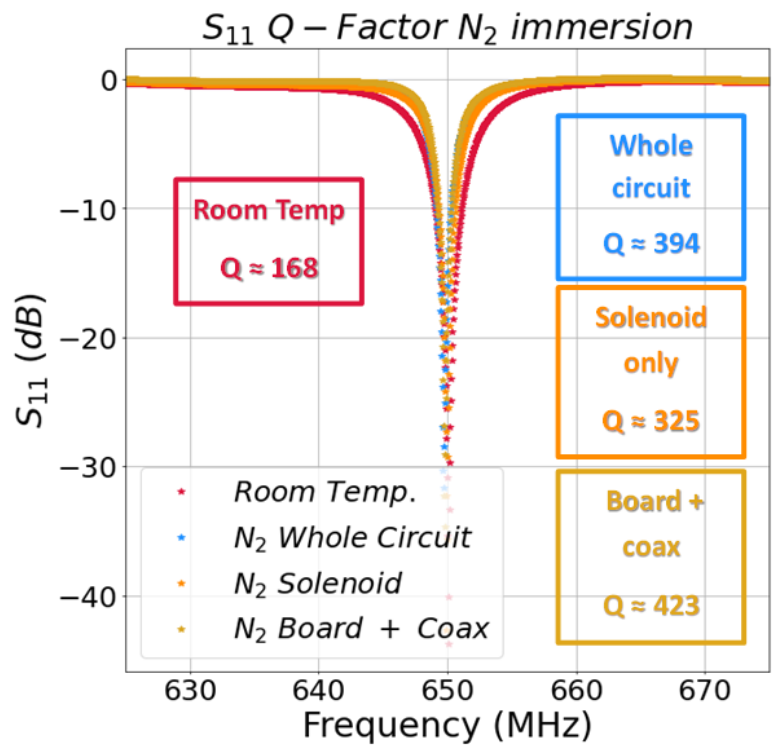

Figure 5: Using a 0.9 mm solenoid, immersing the circuit in liquid $N_2$ results in increased Q factors as characterized by the $S_{11}$ on the bench. The Q-factor enhancement, 168 to 423, corresponds to an SNR increase by at least a factor of 1.6. Further gains from decreased temperatures of the coil may also be realized.

For Johnson noise or thermal noise from the circuit, $R \propto Q^{-1}$. A simple experiment was designed on the bench to investigate if immersing certain regions of the circuit and leaving others in ambient air would result in improved Q factors. Immersing the whole circuit including the solenoid resulted in ΔQ corresponding to an SNR increase of 1.5. Immersing only the 0.9 mm solenoid resulted in ΔQ corresponding to an SNR increase of 1.4. By including 10 cm of the coaxial cable in the immersion, the Q factor increased the most and corresponded to an SNR increase of 1.6. The results of the test are in figure 5.

### 4.2 Quality factor measurements with cryochamber

Several Q-factor measurements were carried out with the cryochamber. The results are summarized in table 1. The initial circuit configuration (circuit 1 in table 1) consisted of 2 variable capacitors and 2 fixed capacitors. Later this design was modified to include only 1 fixed capacitor (circuit 2 in table 1). For a schematic of the two circuit configurations see the supplementary figure 2. Solenoids and surface loops ranging in size from 0.9-4 mm in diameter were investigated on the bench. The final design (surface loop v2 in table 1) demonstrated promising Q improvements corresponding to an SNR improvement of 1.26. The sizes investigated were informed by the calculations in section 2 and results of figure 3. The results in figure 3 agree with the experimental data presented here as a solenoid with a diameter of 4 mm demonstrated no improvements in Q when cooled. The final surface loop design was tested again this time while monitoring the temperature of the outside Kapton insulation of the coil. Once the Kapton reached 6 °C the Q was measured. The Q improved from 34 to 37.4 as the Kapton's initial temperature dropped from 22 to 6°C (final row in table 1).

| Circuit | Geometry | Q (22 °C) | Q (cooled) | $SNR_{N2}/SNR_{RT}$ |
|---|---|---|---|---|
| 1 (no-chamber) | solenoid (N=6, d=0.9 mm) | 168 | 423 | 1.59 |
| 1 (cryochamber) | solenoid (N=10, d=1.5mm) | 116 | 145 | 1.12 |
| 1 (cryochamber) | solenoid (N=10, d=1.5mm) | 126 | 173 | 1.17 |
| 1 (cryochamber) | s. loop v0 (N=1, d=3mm) | 42 | 48 | 1.07 |
| 1 (cryochamber) | s. loop v0 (N=1, d=3mm) | 28 | 37 | 1.15 |
| 1 (cryochamber) | solenoid (N=4, d = 4mm) | 59 | 59 | 1.00 |
| 2 (cryochamber) | s. loop v2 (N=2, d=3mm) | 52 | 83 | 1.26 |
| 2 (cryochamber) | s. loop v2 (N=2, d=3mm) | 34 | *37.4* | 1.05 |
| | | | *italicized red indicates partial cooling* | |

Table 1: Several Quality factor measurements were performed on the bench with and without the cryochamber. The initial test (Circuit 1, no-chamber) simply involved immersing the circuit into a $N_2$ bath. The highest increase in Q was the 2 turn, 3 mm surface loop with ΔQ = 31 and an expected SNR increase of a factor of 1.26.

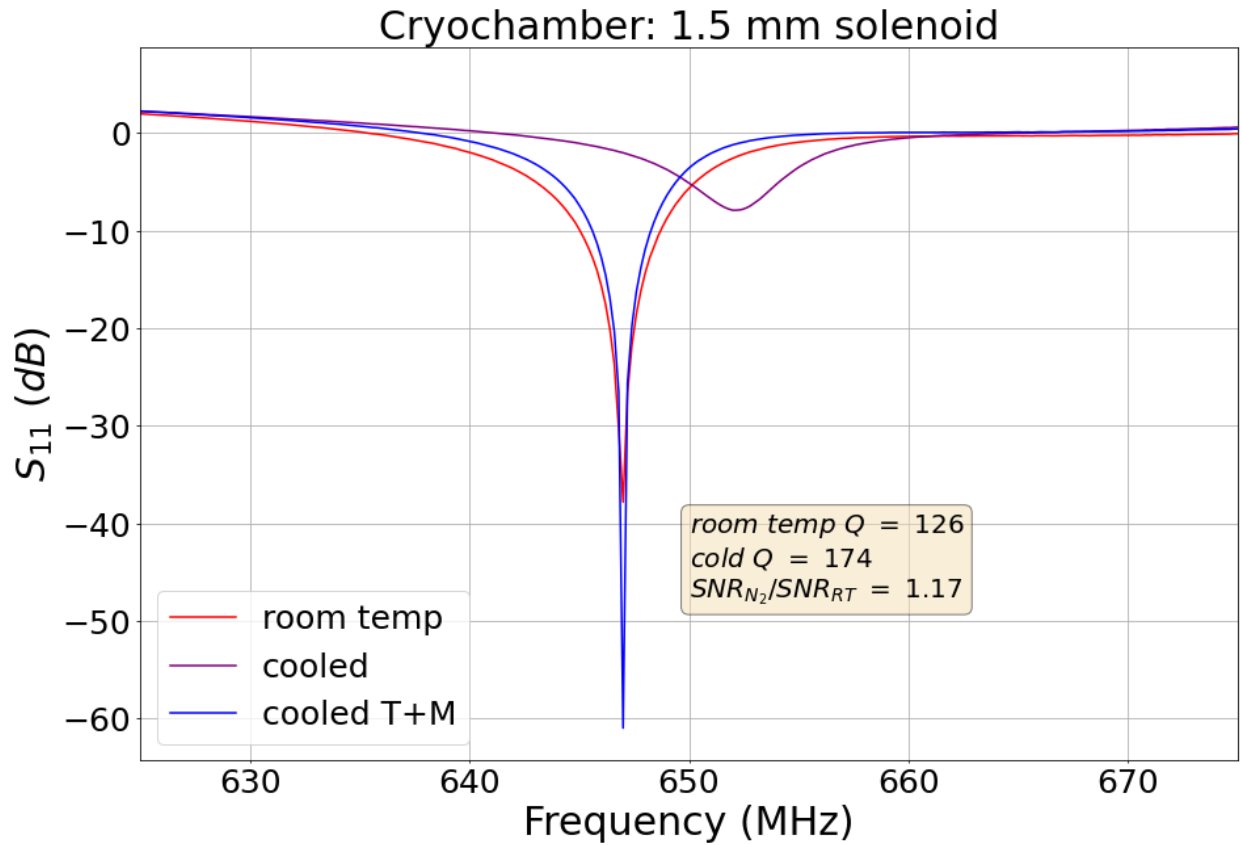

Figure 6: An example of the $S_{11}$ curves before, during, and after cooling. While the system cools the frequency is shifted upwards about 5 MHz. After retuning and matching, the quality factor is improved.

An example Q-factor improvement is shown in figure 6 including the $S_{11}$ curve after the solenoid has cooled down. As the cryochamber is cooled, the resonance shifts upwards ≈ 5 MHz. The circuit must be retuned and rematched back to resonance once cooled.

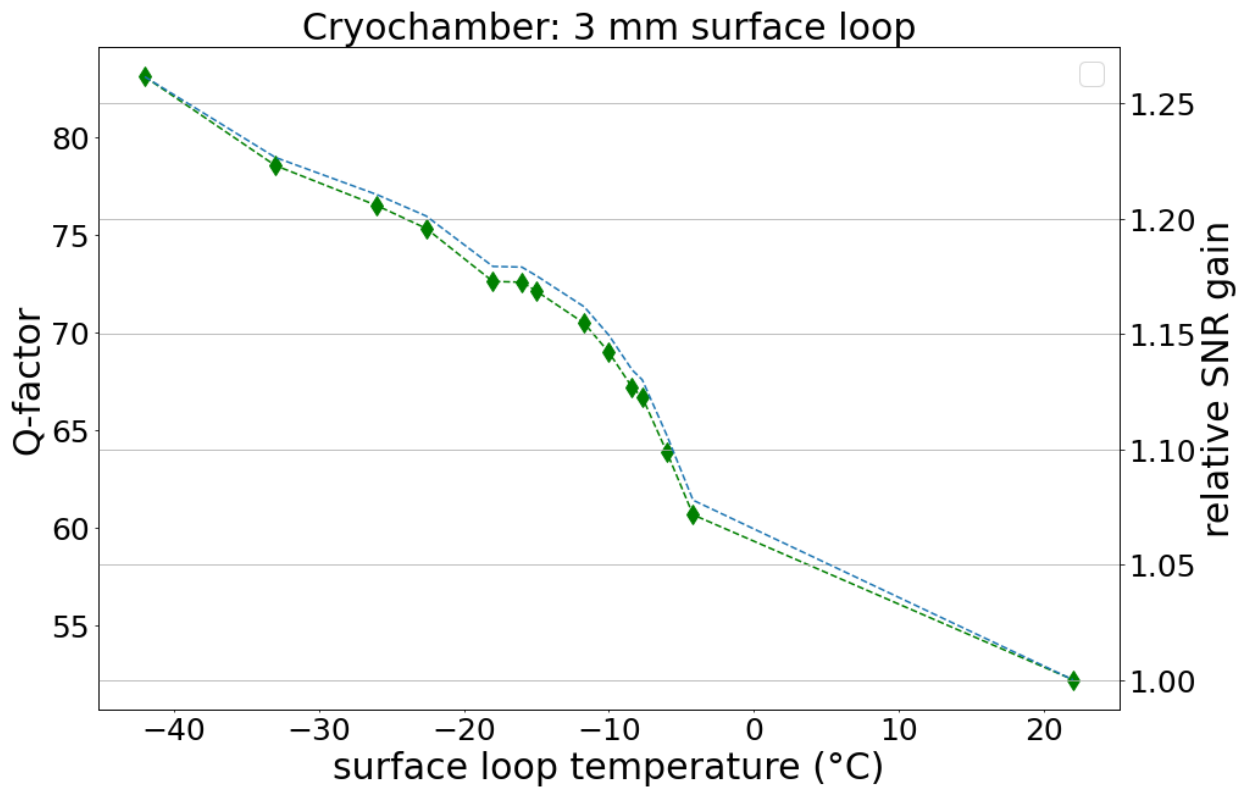

Figure 7: Quality factor measurements were acquired as the coil 2 turn 3 mm surface loop was cooled down. The temperature of the surface loop was measured using a thermocouple. As the surface loop cools, the Q-factor changes significantly with the peak SNR gain being 1.26.

To correlate Q with the temperature of the 3 mm surface loop a thermocouple was placed on the epoxy covered loop while the chamber cooled down. The change in Q and expected increase in SNR is plotted in figure 7. The Q-factor increases from 52 at 22 °C to 83 once the coil reached -42 °C. For each point, the circuit is retuned and matched. The abrupt increase at -5°C to -20°C may be due to the phase changes of the $N_2$ condensing as the internal system cools down.

## 4.3 Imaging Experiments

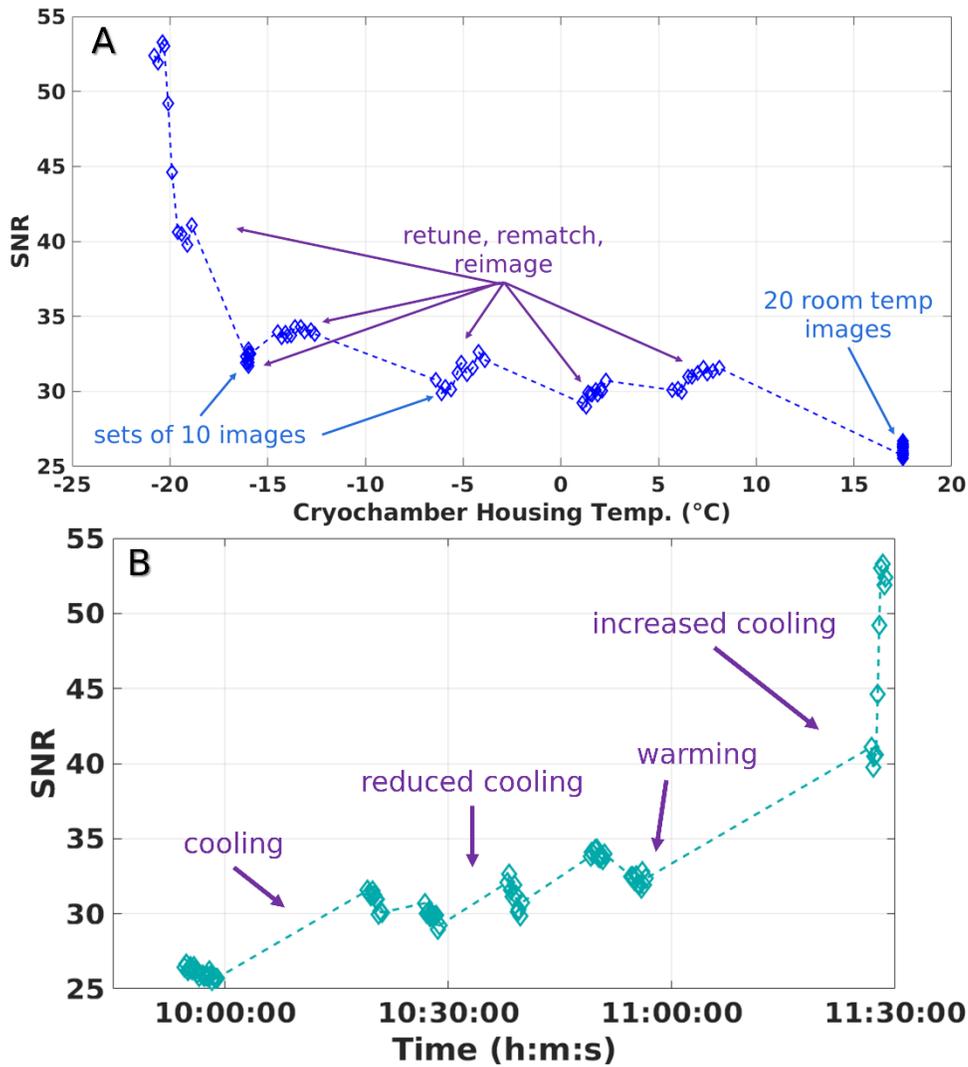

Figure 8: At room temperature, 20 images were acquired to establish a baseline SNR (A). While the cryochamber was cooled down inside the bore, sets of 10 images were acquired to determine any SNR increase from cooling. At each set of images, the $N_2$ was slowed or stopped altogether to ensure the tune and match were adequate (B). At -15 °C, the $N_2$ was stopped to ensure the chamber did not fall significantly below -20°C. The coil was cooled to image as close to the threshold of -20°C as possible, and the final set of 10 images was acquired as the housing reached dropped from -19 to -23°C.

The 20 room temperature images had an average SNR of 26.06 and standard deviation of 0.32. The temperature on the cryochamber housing at these measurements was 17.4 °C. Once the housing dropped 10°C the flow of $N_2$ was interrupted and another set of 10 images was acquired. This set of 10 images had an average SNR of 30.88 and a standard deviation of 0.6. The SNR of each image slowly droops for each image since the $N_2$ is no longer flowing into the chamber (figure 8A). The same pattern can be seen in the next 4 sets of 10 images. The average and standard deviation of the SNR for the next 4 sets of images are as follows: 29.84 ± 0.48, 31.16 ± 0.9, 33.93 ± 0.24, and 32.22 ± 0.35. The time between the $6^{th}$ and $7^{th}$ set of images was much longer than the time between sets 2-6 (figure 8B). This was intentional to avoid damaging the water-cooled gradients by freezing the circulating water within the gradient. Thus, the flow of $N_2$ was slowed between the $6^{th}$ and $7^{th}$ set of images. As soon as the chamber reached -19°C the final set of 10 images was acquired with an average SNR of 46.64 ± 5.86. The final set of 10 images had the following SNR in order of acquisition: 41.09, 39.77, 40.47, 40.6, 44.62, 49.21, 53.03, 53.4, 51.9, 52.4.

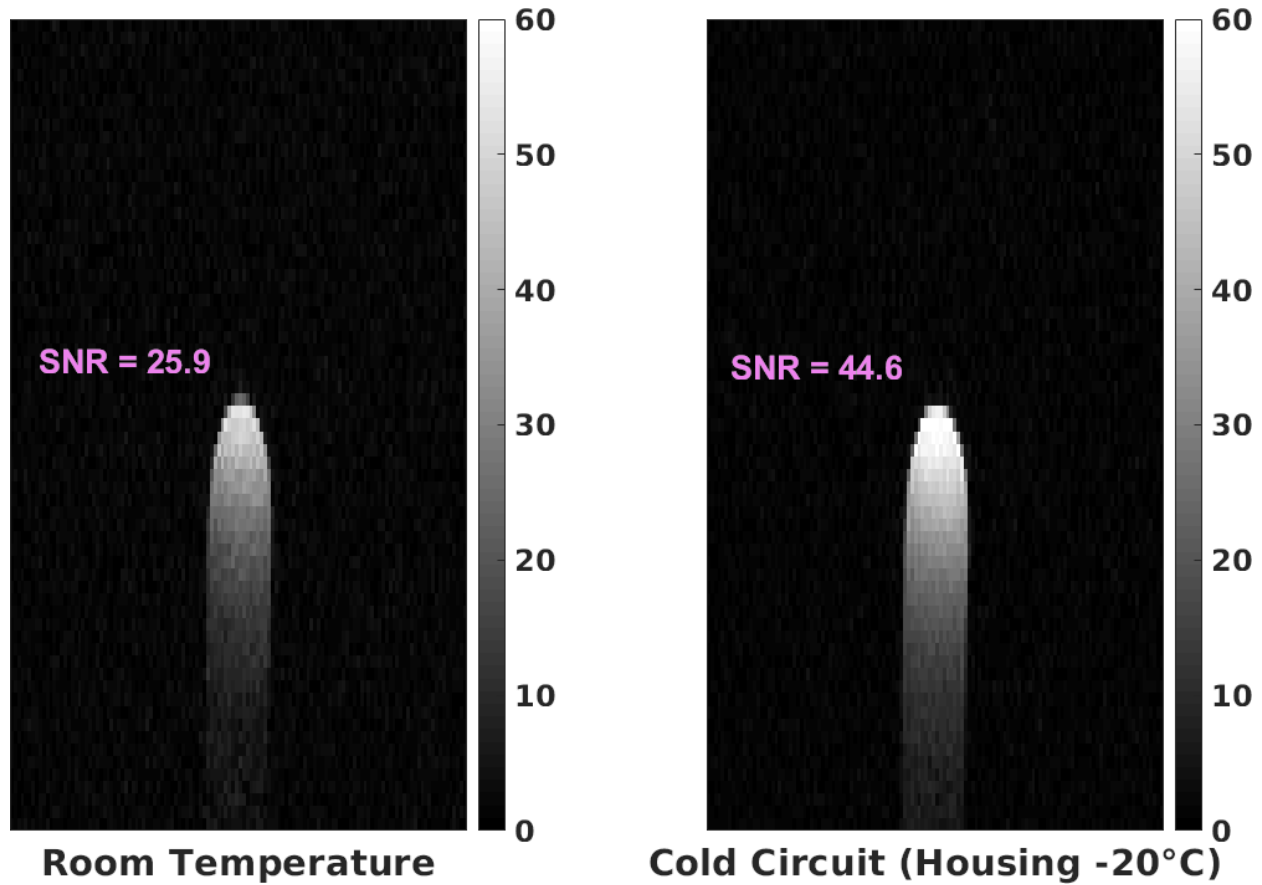

Figure 9: The highest room temperature SNR image (left) compared with the 5[th] image of the final set of images (right). The $SNR_{gain}$ for the two images is 1.72.

In figure 9, a room temperature (17.4°C) image of the $CuSO_4$ phantom is compared with 5[th] image of the final set of 10 images acquired when the housing of the cryochamber reached -20 °C. The SNR increased by a factor of 1.7 for these images. The maximum increase of SNR was 53.28/25.9 or ≈ 2.06.

## 4.4 Effects of cooling on lumped elements

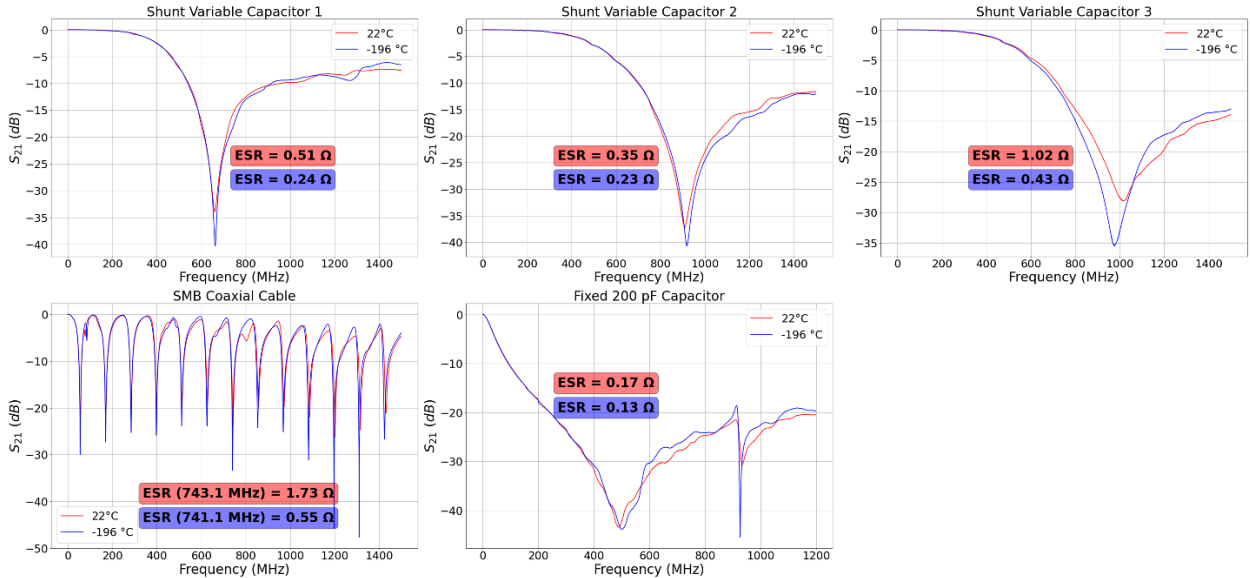

Figure 10: After a full TOSM calibration, five lumped elements were shunted across two coaxial cable connectors. The variable capacitors and coaxial cable saw the largest change in ESR due to cooling. This may be because of the small leads on the variable capacitor contributing to resistance. The periodic peaks for the RG174 coaxial cable correspond to the length of the cable and may be calculated with the frequency distance, the velocity of propagation in the coax, and its dielectric constant.

Figure 10 displays the $S_{21}$ curves for room temperature and cooled lumped elements. The dip of each curve is the self-resonance of each element. Each element's ESR was measured at its self-resonance following appendix 2. For the coaxial cable, the periodic resonances correspond to the length of the cable and the dielectric constant or velocity of propagation in the cable. For the coax, the ESR was measured at around 740 MHz which was the closest dip near 650 MHz. For the first variable capacitor, the $S_{21}$ dropped from -34 to -40 dB corresponding to a decrease in ESR of 0.27 Ω. For the second variable capacitor, the $S_{21}$ dropped from -37 to -41 dB corresponding to a decrease in ESR of 0.12 Ω. For the third variable capacitor, the $S_{21}$ dropped from -28 to -35 dB, corresponding to a decrease in ESR of 0.59 Ω, although the self-resonance was near 1GHz and was not a well-defined peak. For the RG174 coaxial cable, the $S_{21}$ dropped from -24 to -33 dB corresponding to a decrease in ESR of 1.18 Ω. The > 10,000 Q, fixed capacitor's $S_{21}$ peak dropped from -44 to -45 dB corresponding to a much smaller decrease of 0.04 Ω.

The results of the $S_{21}$ and LCR measurements are summarized in supplementary table 1. The variable capacitors ESR at room temperature correspond to the following quality factors: 48, 50, and 16. At -196 °C their quality factors increased to 102, 76 and 37.

## Discussion

We demonstrate that by cooling the tune and match circuit and leaving the microcoil in ambient air it is possible to achieve significantly improved SNR. This is important because it allows more space to insulate the sample and may provoke new cryogenic coil designs using the concept. In general, cooling allows the RF coil designer to trade sensitivity of smaller coils for larger FOVs of cooled coils or decreased scan times. The noise contributions from the circuit are relatively large in comparison to the loop (figure 1), cooling the circuit lowers the total ESR.

Other work has focused on cooling and reducing the noise contributions of the preamplifier [36] and commercial products such as the Bruker Cryoprobe have implemented it [19]. Most work has focused on cooling the coil itself [37–41] and to the authors knowledge none have emphasized the possibility of leaving the loop in ambient air.

The cryochamber (figure 3) is a proof-of-concept designed to demonstrate the improved SNR from cooling of the tune and match circuit (figure 2). Further improvements of the design could include controlled regulation of $N_2$, temperature monitoring of the sample, surface loop, and inner chamber, and control of temperature in the chamber via a proportional-integral-derivative (PID) device (see supplementary figure 4). Temperature stability of the probe is lacking thus prohibiting scan times that exceed a few minutes, a crucial requirement for MR microscopy [13]. The overall design could also be condensed into a smaller chamber to allow more room for insulation between the cryochamber and the inner walls of the bore. Further improvements of the surface loop geometry could be achieved with more sophisticated production techniques including laser milling, thin film deposition, or photolithography of the trace [42]. Although silver has a lower resistivity than copper at room temperature, at 77 K, copper's resistivity is lower [34]. Thus, copper was chosen as the conductor for the trace. As materials with low thermal conductivity and high electrical conductivity are more commonplace and well-understood, RF coils could reap significant SNR benefits as has been demonstrated [43–46].

The fomblin bath is a unique way of controlling the sample temperature while avoiding noise from the flow of water near the FOV (figure 4). Often excised samples are suspended in fomblin or Fluorinert [47] to avoid increased FOVs and added noise from undesired signal. Another method of isolating the sample from extreme temperatures not investigated here is inductively coupling the loop into another passive device that resonates at 650 MHz. This would add further distance of the sample (≈ a coil diameter) from the extreme temperatures.

The cooling of the circuit shifts the frequency higher thus requiring a retune and rematch (fig 5 and 6). The shift upwards is most likely from a decrease in resistance and capacitance in the system. The decrease in capacitance could be due to the dielectric constant decreasing rapidly as the materials in the circuit go through phase changes [48]. The frequency shift is evidence that the impedance of the circuit is changing. The drastic phase change of $N_2$ gas to liquid may also explain the abrupt SNR changes in figure 7 and 8. In figure 7, when the surface loop temperature passes 0 °C, the internal chamber could be filling with liquid $N_2$. In figure 8 during the final set of 10 images, the SNR quickly improves from 40 to 53. The 30-minute gap between the last two data sets in figure 8B was purely out of caution. We did not want to exceed -25 °C and risk freezing the circulating water within the 6 cm gradient coil. During the 30-minute gap, the transfer line had inevitably heated to above freezing (visible condensation) although the cryochamber remained below -11 °C. To approach -20 °C, the dewar was reopened and $N_2$ slowly re-cooled the transfer line. The re-cooling took ≈ 15 minutes. The final images were acquired as the chamber cooled from -19 to -23 °C yielding the final image in figure 9.

In figure 10, the $S_{21}$ measurement is intended to show that cooling decreases the circuit components ESR. The exact values may not be accurate. For example, at 500 MHz the 200-pF fixed capacitor is reported by the vendor to have an ESR of 0.05 Ω. This is 3x less than the ESR reported in figure 10. However, the measurement is not intended to provide exact values of the ESR, it is only intended to show the temperature decreases the ESR of the components. The exact ESR values would need to be measured with more precise and standardized resonant line methods [31].

The difference in predicted $SNR_{gain}$ in figure 2 (≈ 1.5) vs the measured $SNR_{gain}$ in figure 9 (≈ 2) could be due to a few reasons. In the calculated plots, the coaxial cable was not included since the data sheets do not include the loss of the cables as a function of temperature. Similarly, the capacitors' Q = 2000 for room temperature and the cooled circuit. This is inaccurate as the ESR of the variable capacitors and thus Q depends on temperature (see figure 10).

## Conclusion

By cooling the tune and match circuit and leaving the surface loop in ambient air, SNR was improved by a factor of 2. The results are significant because it allows for more space to insulate the sample from the extreme temperatures. A cryochamber was designed to cool the tune and match circuit within the bore without damaging the gradient system. A unique fomblin bath was designed to insulate the sample from extreme temperatures without introducing further noise from circulating water.

## Acknowledgments

This work was supported by the Chan-Zuckerberg Initiative for Deep Tissue Imaging.
The authors would also like to acknowledge Sai Abitha Srinivas for helpful discussions, Franz Baudenbacher for insightful cryogenic advice, Dan Colvin for experimental execution, and John Keiley for discussion regarding the temperature tolerance of the water-cooled gradient coil.

## Appendix 1

Beginning at the front of the chamber, the surface loop (supplementary figure 3) is soldered onto leads protruding from the cryochamber and sealed with epoxy (figure 3B). The loop was placed ≈ 1.5 mm away [49] from the sample.

The cryochamber is a 3.19 cm OD, 2.54 cm ID garolite tube threaded on each end so that custom garolite caps are screwed in with a watertight seal. Each thread was wrapped tightly with 3-5 layers of Polytetrafluoroethylene (PTFE) tape and were fitted with low temperature O-rings. The threaded front cap of the cryochamber was hollowed out so that the small tune and match circuit could fit tightly inside. The width of the garolite wall between the surface loop and the circuit where $N_2$ flows freely was ≈ 3 mm. The tune and match circuit was as simple as possible to ensure it fit within the < 2.54 cm diameter cap. The hollow slot was machined to ≈ 20 mm to fit the circuit snugly. The circuit consisted of a variable series matching capacitor, a variable parallel tuning capacitor, and a fixed series capacitor (supplementary figure 2) to ensure the circuit was balanced.

The dual-walled chamber consisted of the 2.54 cm ID garolite tube surrounded by ultra-thin low temperature aerogel insulation. The insulation was then surrounded by the 2$^{nd}$ wall which consists of ultra-low temperature garolite tube with OD 5.08 cm and ID of 4.45 cm. The front of the dual walled chamber has a larger cap that threads into the outermost tube and surrounds the innermost tube with an opening for the surface loop leads (figure 3 B,C). The back of the dual walled chamber had another garolite cap that threaded into the outermost tube. The two front and back caps were necessary to maintain the position of the internal tube which was sitting on layer of aerogel insulation. The threaded external back cap also functioned as a fixed point for the o-ring compression mechanism (figure 3D).

The internal cap threaded into the cryochamber had 5 inputs; $N_2$ input tube, exhaust tube, 2 tune rods, and the coaxial cable. The $N_2$ input tube was threaded and sealed with PTFE tape. The coax and exhaust tube were sealed with an external layer of epoxy. To allow the tune rods to rotate while maintaining a liquid $N_2$ tight seal, low temperature vacuum grease and o-rings were compressed by a custom garolite piece against the wall of the internal cap. The o-ring compression piece was tightened with screws into the external cap. To maintain the seal, the whole chamber was cooled down to liquid $N_2$ temperatures and the screws were retightened to ensure any shrinkage from lower temperatures was accounted for. $N_2$ input flowed directly from a 240 L dewar with a 6.1 m vacuum jacket extension. The extension was sealed to a barb end using PTFE tape. A 10 cm plastic tube connected the barb to the $N_2$ input at the back panel of the cryochamber. The tube was clamped to the barb and $N_2$ input tube for a tight seal. Surrounding the external cap and o-ring compressor was another 5.08 cm OD low temperature garolite tube. The tube followed the input and exhaust tubes, coax, and tune rods as structural support and acted as another layer of insulation surrounding the input tube (figure 3E). A small amount of aerogel insulative material was placed inside the housing although most of its volume was ambient air. The remaining space between the housing and wall of the bore (6cm diameter bore) was filled with aerogel insulation, a small layer of heat shrink, and Kapton tape (figure 3F). The materials list and CAD files for the cryochamber may be obtained from the supplementary material.

## Appendix 2

Data sheets of the capacitors do not contain Q measurements at varying temperatures. A network analyzer (NA) was used to assess the effect of temperature on the ESR of the tune and match circuit. The ESR measurement may be performed with some caveats. Mainly the measurement only accurate at self-resonances below 1 GHz to avoid confounding influences of parasitic high frequency effects. Placing circuit elements as shunted circuits (i.e. parallel) at the ends of coaxial cables is the most accurate method of measuring ESR with a network analyzer [50]. Each element was placed in parallel with two BNC connectors after a full through open short match calibration (TOSM) and $S_{21}$ was measured. The ESR at the self-resonance of the element may then be calculated using the following equation, derived from simple 2-port systems.

$$ESR\ (S_{21}) = \frac{Z_0}{2\left(\dfrac{1}{10^{S_{21}/20}} - 1\right)}$$

where $Z_0$ is the characteristic impedance of the transmission line which is 50 Ω. 3 variable capacitors, an open 87 cm length of RG174, and a fixed 200 pF capacitor were placed in parallel between the BNC connectors. The lumped element was then carefully placed over the $N_2$ bath. The $S_{21}$ curve was acquired. The element was then lowered into the $N_2$ bath and the $S_{21}$ measurement was acquired again.

Using an LCR meter, ESR measurements at 100kHz were done on the following circuit elements: 3 variable capacitors, the surface loop, a fixed 39 pF capacitor, shorted RG174 coaxial cable, and a shorted SMB to PCB connector. The measurement was done by carefully placing each element between the LCR probe ends. 10-15 ESR measurements were taken at room temperature. The element was then gently dipped into an $N_2$ bath, and 10-15 measurements of ESR were taken. Before measuring the subsequent elements, the probe was allowed to return to ambient temperatures.

## Supplementary material

| Lumped Elements | 100 kHz (22 °C) | 100 kHz (-196 °C) | $f_{res}$ (22 °C) | $f_{res}$ (-196 °C) |
|---|---|---|---|---|
| surface loop (mΩ) | 52 | 44 | ~ | ~ |
| 0.3-10 pF variable cap 1 (Ω) | 718 | 474 | 0.51 | 0.24 |
| variable cap 2 (Ω) | 731 | 430 | 0.35 | 0.23 |
| variable cap 3 (Ω) | 638 | 495 | 1.02 | 0.43 |
| 39 pF fixed cap (Ω) | 271 | 137 | ~ | ~ |
| shorted SMB connector (mΩ) | 161 | 136 | ~ | ~ |
| shorted SMB coax (mΩ) | 217 | 174 | (open) 1.73 Ω | (open) 0.55 Ω |

Supplemental table 1: LCR meter measurements at 100 kHz were obtained at room and liquid $N_2$ temperatures for select elements of the circuitry in the cryochamber. $S_{21}$ of self-resonating elements ($f_{res}$) are also included. ~ indicates that the element did not resonate when shunted across the two-port setup described in 3.4.2.

Supplementary figure 1: For a solenoid, the dielectric losses dominate the magnetic losses. As the number of turns increases, the number of electric field lines passing through the sample increase, drastically increasing the losses.

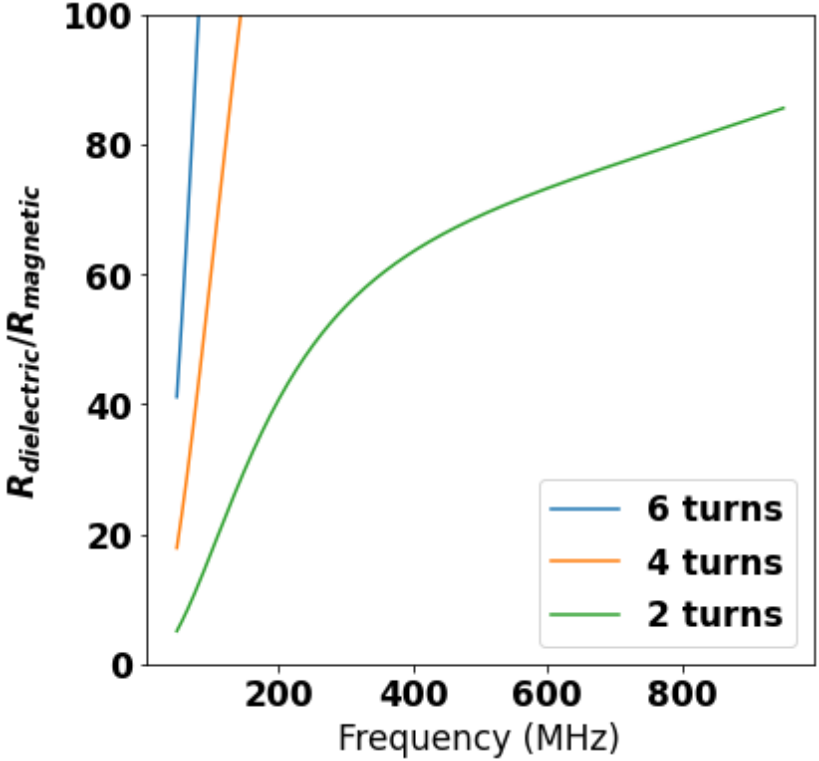

Supplementary figure 2: The two circuit schematics used in the cryochamber. The circuits shared the same width of 15.2 mm. The ground plane of circuit 2 was the 2nd side of a double sided PCB board.

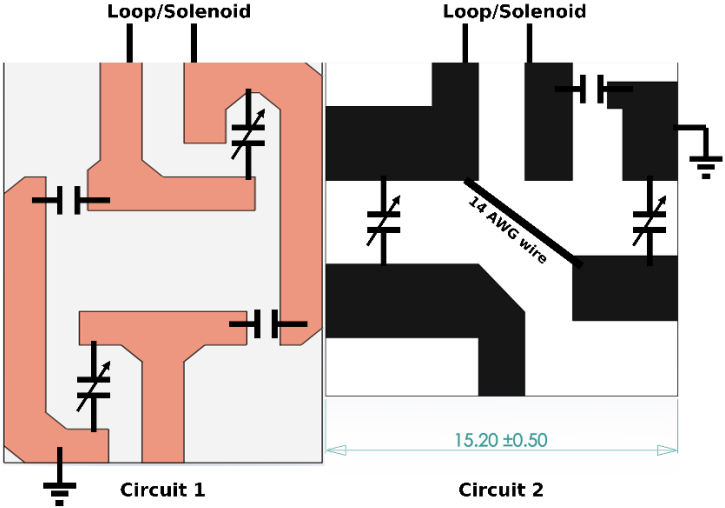

Supplementary figure 3: The 2 turn, 3mm surface loop with corresponding dimensions in mm.

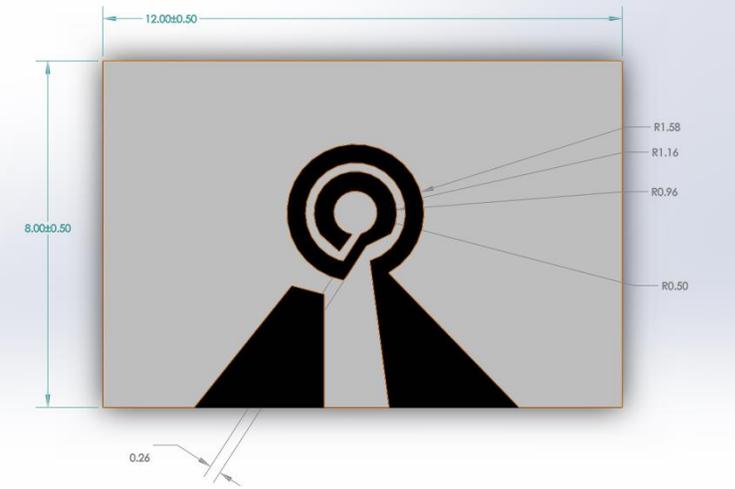

Supplementary figure 4:

A PID controller was designed for temperature control. The setup consisted of a heater, a 5V relay switch, and a programmable Raspberry Pi. A python script utilizing a PID library was run to control the temperature of the thermocouple. The thermocouple temperature data was read on a laptop and transferred via scp to the raspberry pi for PID calculations. The scripts and transfer protocol are available at https://github.com/benjhardy/Coil_vs_Sample_Noise_MRI.

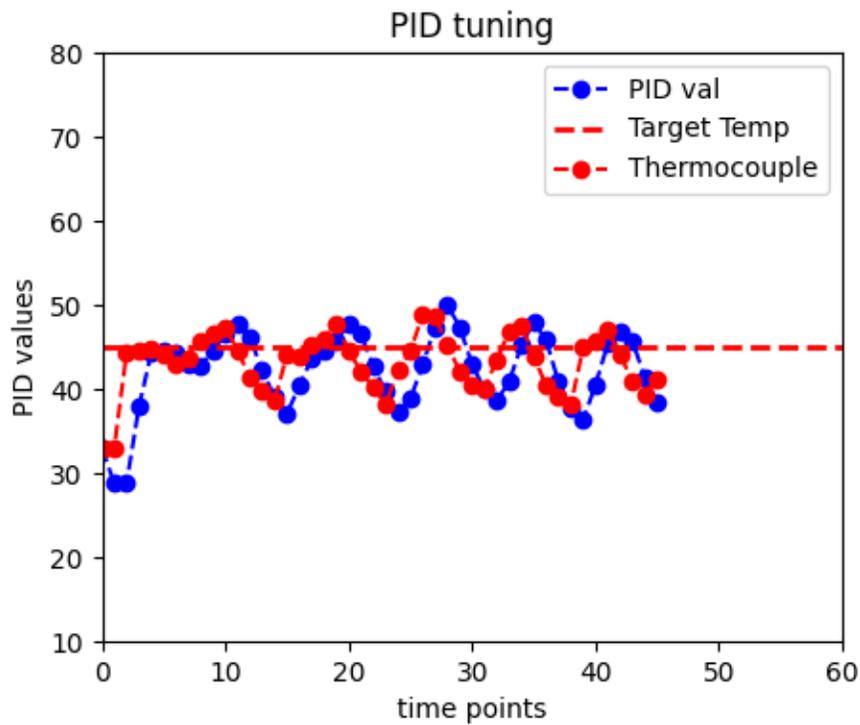